# Size dependence of the $T_c$ and the superconducting energy gap in nanocrystalline thin films of Nb

Sangita Bose, Pratap Raychaudhuri, Rajarshi Banerjee*, Parinda Vasa and Pushan Ayyub

Department of Condensed Matter Physics and Material Science,

Tata Institute of Fundamental Research, Mumbai 400005, India

*Center for the Accelerated Maturation of Materials, Department of Materials Science

and Engineering, The Ohio State University, Columbus, Ohio 43210

In nanocrystalline Nb films, the superconducting $T_c$ decreases with a reduction in the average particle size below ≈20nm. We correlate the decrease in $T_c$ with a reduction in the superconducting energy gap measured by point contact spectroscopy. Consistent with the Anderson criterion, no superconducting transition was observed for sizes below 8 nm. We show that the size-dependence of the superconducting properties in this intermediate coupling Type II superconductor is governed by changes in the electronic density of states rather than by phonon softening.

Despite several decades of active research,[1, 2, 3, 4, 5, 6, 7, 8, 9, 10] a satisfactory understanding of the modification of superconductivity at reduced length scales is yet to be achieved. Changes in superconducting properties are expected when the effective size of the superconductor is reduced below its characteristic length scales, such as the London penetration depth $\lambda_L(T)$ or the coherence length $\xi(T)$. Since long-range order cannot persist down to the zero dimensional limit, it is interesting to probe how the superconducting properties evolve below these length scales. A complete destabilization of superconducting order is expected when the system size is so small that the electronic level spacing becomes larger than the bulk superconducting energy gap ($\Delta$). Thus, the so-called Anderson criterion,[2] suggests that, in principle, superconductivity may persist at length scales far below $\xi$. Experimentally, the studies of superconducting properties of mesoscopic systems fall broadly in two categories. The first category of experiments deals with isolated grains or compacted powders. In weak and intermediate coupling type I superconductors such as Al[1,3,4] In[1,4] and Sn[4,5], an enhancement in $T_c$ has been reported for particle sizes below $\approx$100nm. In the strong coupling superconductor, Pb[1,4,6] most of the early reports show no change in $T_c$ down to a few nm. However, in a recent work Li et al[7] reported that in isolated grains of Pb, $T_c$ decreases with the grain size down to a critical size, below which there is complete loss of superconductivity. In an attempt to explain these observations, a number of mechanisms have been invoked, which take account of the discretization of the electronic energy levels[2,21] and softening of the surface phonon modes.[11, 17, 18] Based on these calculations, both an increase as well as a decrease in $T_c$ have been predicted. The second category of experiments was carried out on quench condensed, disordered superconducting thin films (such as Sn, Pb or In) evaporated on substrates kept at liquid helium temperature.[8,10,11] These films consist of small superconducting islands with size varying between 100-200nm and connected through a two-dimensional network of Josephson junctions. Though an accurate measurement of the particle size in these films is difficult, a gradual destruction of superconductivity is observed with increasing sheet resistance when the film thickness is reduced to less than ~ 5 nm. The loss of superconductivity in these films has been attributed to rapid phase fluctuations between the grains in the two-dimensional network though the individual grains retain their superconducting properties.[10]

In this paper we investigate the evolution of $T_c$ and the superconducting energy gap ($\Delta$) with decreasing particle size in nanocrystalline thin films of Nb. For Nb, the value of the electron-phonon coupling strength, $2\Delta/k_B T_c \approx 3.8$, falls between In ($2\Delta/k_B T_c \approx 3.6$) and Pb ($2\Delta/k_B T_c \approx 4.5$).[12]. Our experiments were performed on nanocrystalline thin films with average grain sizes ($D$) in the 5nm-50nm range, synthesized by high pressure magnetron sputtering. All films with $D > 20$nm show bulk superconductivity with $T_c \approx 9.2$K, while there is a gradual suppression of $T_c$ between 20nm and 8nm. No superconducting transition was detected for $D < 8$nm, consistent with the Anderson criterion. Clearly, there is no size-dependent enhancement in $T_c$ in nano-Nb. Finally, we demonstrate a direct correlation between the measured values of $\Delta$ and $T_c$ in nano-Nb, which was found to remain in the weak coupling BCS regime down to the lowest particle size studied. This suggests that the evolution of superconducting properties with particle size in Nb is distinct from the other weak and strong coupling superconductors studied so far.

Nanocrystalline thin films of Nb were deposited using high-pressure magnetron sputtering in a custom-built UHV chamber (base pressure prior to sputtering ~5x10$^{-8}$ Torr) using an elemental Nb target of 99.99% purity. All depositions were carried out on oxidized Si substrates, which had a 200nm thick, thermally grown amorphous oxide layer on the surface. The particle size was varied in the 5-50nm range by controlling process parameters such as the sputtering gas (Ar) pressure, the power, the deposition time and the substrate temperature. The average particle size of the as-deposited Nb films was determined using x-ray diffraction line broadening (Scherrer technique) and transmission electron microscopy, and the film thickness was in the 0.2-1μm range. Prior to introduction into the sputtering chamber, the Ar gas was purified by passing it through a gettering furnace with a Ti sponge maintained at 1073K. Details of the sample deposition and characterization are described elsewhere.[13]

The sample thus consists of a dense aggregate of nanoparticles that can be visualized as a network of weakly connected superconducting grains. The weak link Josephson junction character of the grain boundaries was established from a study of the current-voltage characteristics. To this extent the nanocrystalline films are similar to the quench condensed films,[8,10] but the similarity ends here since the film thickness does not affect superconducting properties in the essentially 3D nano-Nb films. Magnetization and transport measurements were carried out using a Quantum Design MPMS SQUID magnetometer and a liquid-He cryostat respectively. The superconducting energy gap (Δ) was measured using a custom-designed point contact spectroscopy setup where the temperature could be controlled down to 1.6K. A mechanically cut Pt-Ir tip was brought in contact with the film using a differential screw technique and the conductance spectra (dI/dV vs. V) of the resulting contact was analyzed to determine the value of $\Delta$.

Figures 1(a) and (b) show the temperature dependence of the magnetization and resistivity for the nano-Nb samples with different particle size. The $T_c$s obtained from both measurements coincide for all the samples (Fig. 1(c), and we observe a monotonic decrease in the $T_c$ from 9.4K to 4.7K between 20nm and 8nm. Below 8nm, no superconducting transition was observed down to the lowest measurable temperature of 1.76K. Figure 1(c) also shows that there is a steady increase in the lattice parameter with decreasing size, the origin of which has been discussed elsewhere.[13] Since our measurements were made on nanocrystalline films with closely packed grains, it is important to establish that the electrons are localized within the grains, before ascribing the observed variation of $T_c$ to finite size effects. The superconducting grains need to be weakly coupled for the film as whole to behave as a disordered Josephson junction network with a well-defined, macroscopic $T_c$. Earlier studies on quench condensed films have shown that such a weakly coupled disordered Josephson junction network[14] displays a pronounced hysteresis in the current voltage (I-V) characteristics when the current is swept up and down across the critical current. Figure 2 shows the I-V characteristics of the Nb films with different particle size (and $T_c$). Apart from the film with the largest size ($D$ = 50nm, $T_c$ = 9.2K), all others show a pronounced hysteresis with increasing and decreasing current, confirming the weakly coupled nature of the grains. The reduction in $T_c$ can therefore safely be attributed to size effects.

To determine the evolution of the superconducting energy gap with particle size, we carried out point contact spectroscopic measurements on the films by contacting then with mechanically cut Pt-Ir tips at low temperature. Several contacts were made for each film and the conductance vs voltage ($G = dI/dV$ vs. $V$) characteristics of the junction recorded. Figure 3(a)-(d) show the representative $G$-$V$ spectra for contacts on four films with different particle size. The value of $\Delta$ was determined by fitting the spectra with the Blonder-Tinkham-Klapwijk (BTK) theory[15] with $\Delta$, $Z$ (barrier parameter) and $\Gamma$ (broadening parameter[16]) as fitting parameters. $\Delta$ varies from its bulk value, $\Delta(0) \approx 1.60$ meV for the film with $D = 50$nm to $\Delta(0) \approx 0.90$ meV for $D = 10.4$nm ($T_c \sim 5.9$K). The variation of $\Delta$ with the observed $T_c$ is shown in Figure 4(a). The direct correlation between $T_c$ and $\Delta$ confirms that the destruction of superconductivity in our films is an intrinsic property of the grains and not caused by rapid phase fluctuation between adjacent grains as observed in quench condensed films in which $\Delta$ remains unchanged.[10] Further, the linear relation between $\Delta$ and $T_c$ confirms that Nb remains in the intermediate coupling limit even for the smallest size. This is also confirmed by the observed temperature variation of $\Delta$, which can be fitted well (even for the smallest size) with the behavior expected from the weak coupling BCS theory (figure 4(b)).

To understand these results, we now briefly outline the two mechanisms usually invoked to account for the size dependence of $T_c$, namely, the softening of the phonon spectrum[11, 17] and the change in the electronic density of states due to discretization of the electronic energy bands. The first mechanism assumes that individual grains consist of a surface shell in which the phonon frequency is reduced from its bulk value. Using the McMillan equation,[18] which relates the phonon frequency to the electron phonon coupling constant, one can show that this results in an increase in the effective electron phonon coupling, and hence also in the $T_c$. This is widely believed to be the reason for the $T_c$ enhancement observed in weak coupling superconductors.[3, 11] In strong coupling superconductors or for very small particle sizes this effect is offset by an increase in the low frequency phonon cutoff arising from quantization of the phonon wave-vector, and effectively suppressing the $T_c$.[19, 20] Note that in this case, the change in $T_c$ caused by the effective change in the electron phonon coupling constant, should also produce a change in $2\Delta/k_BT_c$. In the particular case of strong coupling superconductors, where $T_c$ decreases with decreasing $D$, $\Delta$ is nevertheless expected to show a small increase. This is clearly incompatible with our observations.

The second mechanism is based on changes in the electronic density of states due to quantization of the electron wave vector **q.** Strongin et al[21] theoretically explored this effect in an attempt to explain the destruction of superconductivity in very small superconducting particles. Solving the discrete version of the BCS equation, the quantization of electronic states gives a natural suppression of $\Delta$ through a modification of the electronic density states at the Fermi level. Any variation in the $T_c$ arising from this mechanism should be associated with a proportional variation[20] in $\Delta$. The linear variation of $\Delta$ with $T_c$ in our films strongly indicates that the second mechanism plays a dominant role in nano-Nb while the effect of softening of the phonon spectrum is almost negligible[22].

It is pertinent to ask why the softening of the phonon spectrum so ubiquitously claimed to play a dominant role in superconducting nanoparticles is not observed in Nb. While we do not yet have a complete explanation, a few points are worth considering. The softening of the surface phonon modes has been typically demonstrated though a molecular dynamic simulation for particles with a thickness of a few monolayers.[17] This effect could be compensated if the surface phonon spectrum is qualitatively different from the bulk, i.e., if the surface stabilizes in a crystallographic structure different from the bulk. This is indeed a possibility in Nb, in view of the appreciable lattice expansion ($\approx$6%) exhibited at small particle sizes (Fig. 1(c)). This is in sharp contrast with In, Sn or Al where the change in lattice parameter is less than 0.2%. We also point out that an unambiguous proof of the involvement of surface phonon softening in the size dependence of $T_c$ requires a direct measurement of $\Delta$ and $T_c$ (such as reported here) to find out if $2\Delta/k_B T_c$ varies with particle size. Such measurements would be of great interest for superconductors where this mechanism has been speculated to play a significant role.

In summary, we have reported a detailed investigation of the evolution of superconductivity with particle size in nanocrystalline films of Nb. We observe that the electron phonon coupling is not significantly affected by a decrease in the particle size and the superconductor remains in the intermediate coupling limit down to 8nm. Our results strongly indicate that the size dependence of superconductivity in Nb is primarily governed by the changes in the electronic density of states as opposed to changes in the electron-phonon coupling due to surface effects, as suggested earlier in In, Sn and Al. However, even in those systems it would be instructive to confirm the validity of the surface phonon softening theory through a detailed measurement of $2\Delta/k_BT_c$ to estimate the actual change in electron-phonon coupling strength with decreasing particle size.

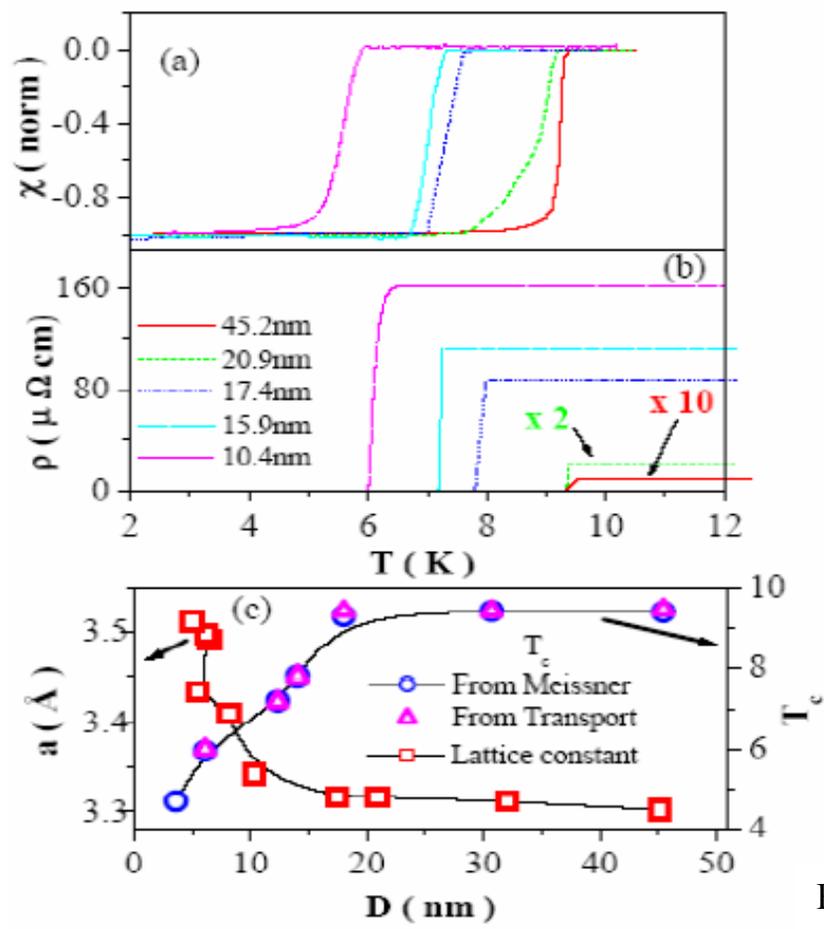

Figure 1

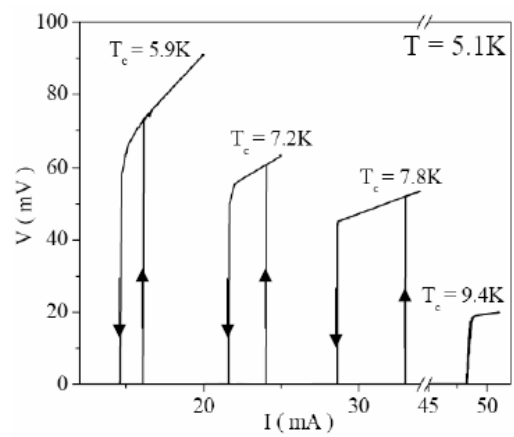

Figure 2

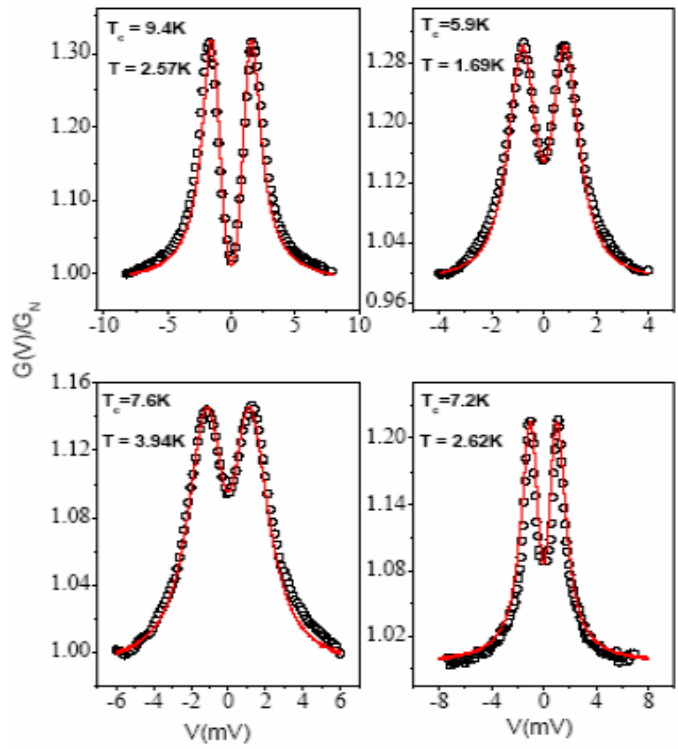

Figure 3

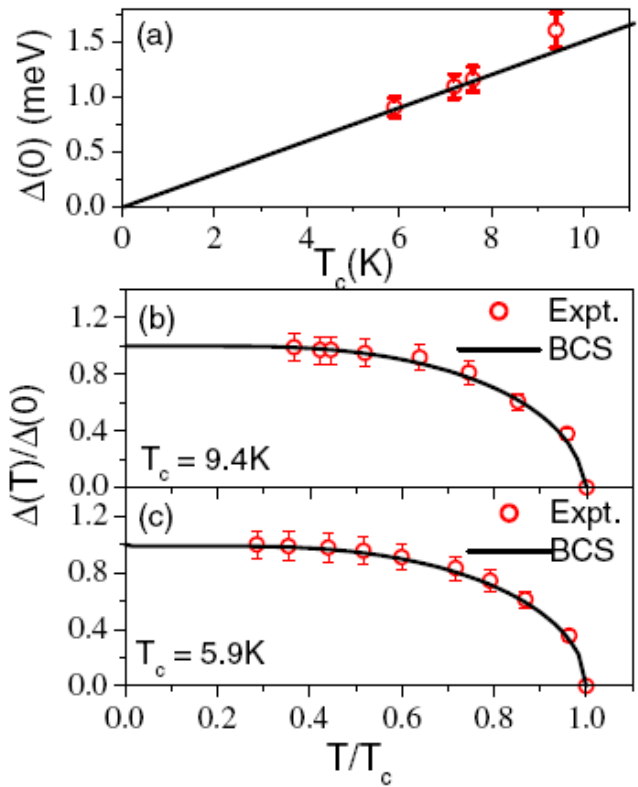

Figure 4

Fig 1: Temperature dependence of (a) magnetization, and (b) resistivity in nanocrystalline thin films of Nb with different average particle size. (c) Size dependence of the superconducting transition temperature ($T_c$) obtained from magnetization (open circle) and resistivity (open triangles) measurements. The figure also shows the size dependence of the lattice parameter (open squares) in nano-Nb. (The solid lines are intended to guide the eye.)

Fig 2: I-V characteristics of the nanocrystalline Nb samples with $T_c$ = 9.4K, 7.6K, 7.2K, and 5.9K measured at 5.1K.

Fig 3(a-d): Point Contact Andreev Reflection spectra for nanocrystalline Nb samples with $T_c$ = 9.4K, 7.6K, 7.2K, and 5.9K. The dotted curve shows the data while the solid line is the BTK fit.

Fig 4(a): Variation of superconducting gap $\Delta(0)$ with the $T_c$, and (b) Temperature dependence of $\Delta(0)$ with the reduced temperature for two different samples of nano-Nb, with $T_c$ = 5.9K, and $T_c$ = 9.4K.

[22] By comparing their data on Pb films Strongin et al. concluded that this effect alone couldn't account for suppression in $T_c$. Their results however were on quench-condensed films. It has been later demonstrated that in 2-D disordered films of this kind the suppression of superconductivty is caused by rapid phase fluctuations (ref. 10), an effect which has been neglected in Strongin's calculation.